\documentclass[apj]{emulateapj}
\usepackage{graphics}
\usepackage{upgreek}
\shorttitle{Evolution of the PPDs}
\shortauthors{Khajenabi, Kazrani, Shadmehri}

\begin{document}

\title{An analytical model for the evolution of the protoplanetary discs}

\author{Fazeleh Khajenabi, Kimia Kazrani, Mohsen Shadmehri }
\affil{Department of Physics, Faculty of Sciences, Golestan University, Gorgan 49138-15739, Iran \\ f.khajenabi@gu.ac.ir}

\begin{abstract}
We obtain a new set of analytical solutions for the evolution of a self-gravitating accretion disc by holding  the Toomre parameter  close to its threshold, and obtaining the stress parameter from the cooling rate. In agreement with the previous numerical solutions, furthermore, the accretion rate is assumed to be independent of the disc radius. Extreme situations where the entire disc is either optically thick or optically thin are studied independently and the obtained solutions can be used for exploring the early or the final phases of a protoplanetary disc evolution. Our solutions exhibit decay of the accretion rate as a power-law function of the age of the system with the exponent -0.75 and -1.04 for optically thick and thin cases, respectively.  Our calculations permit us to explore evolution of the snow line analytically. Location of the snow line in the optically thick regime evolves as a  power-law function of time with the exponent -0.16, however, when the disc is optically thin, location of the snow line as a function of time with the exponent -0.7 has a stronger dependence on time. It means that in an optically thin disc inward migration of the snow line is faster than an optically thick disc. 
\end{abstract}


\section{Introduction}

Although most of the analytical models for the structure of protoplanetary disks (PPDs) are time-independent \cite[e.g.,][]{rafikov2009}, further studies have revealed that a PPD undergoes significant changes during its lifetime and its physical quantities are actually functions of time.   \cite{step98b} advanced the idea that a better physical insight about PPDs is gained by examining their evolutionary path, and he arrived to this key finding that a conventional disc model  is more consistent with the observations, instead of the layered accretion model. Since the typical lifetime of a PPD is estimated to be between $10^6$  and $10^7$ years, it is reasonable to expect that any possible changes in the structure of a PPD should occur less than this typical time-scale. \cite{hartmann98} examined time-dependent behavior of the accretion rate in T Tauri stars using similarity solutions with the disc viscosity as a power-law function of the radial distance. They found a clear decline of the accretion rate with the age of the system.

The usual treatment for examining time-dependent behavior of a thin accretion disc within the framework of the standard model \citep{SSmodel} is based on the conservation of mass and angular momentum which eventually leads to a diffusion-type partial differential equation for the surface density.  Solutions of this equation, irrespective of the true sources of the angular momentum transport, have been investigated in most of the previous analytical models for the evolution of thin accretion discs. The key input parameter is the proposed viscosity, however, various functional forms have been suggested for the viscosity depending on the mechanisms for generating turbulence. Most of the previous studies on the evolution of a thin discs are limited to the cases where viscosity can be written as a power-law function of the radial distance and the surface density.

When the viscosity is a power-law function of the radial distance, the resulting diffusion-type equation for the evolution of the surface density becomes linear and the standard methods for solving linear partial differential equations, such as Green function, can be used to obtain  analytical solutions \citep[e.g.,][]{lynden74,pringle91,tanaka2011,Lipu15}. Depending on the imposed inner or outer boundary conditions and  the size of the disc, however, different types of solutions have been obtained with a common feature that the accretion rate decays with age of the system. \cite{lynden74}, for instance, found a power-law decay for the accretion rate, i.e. $\dot{M} \propto t^{-\eta}$ where $1<\eta <2$. Their self-similar solutions  are appropriate for the discs with infinite radial extension. However, \cite{king98} in their study to explain the light curves of soft X-ray transients showed that disc irradiation by the central source causes the accretion rate decay to be exponential.  In T Tauri stars, \cite{hartmann98} estimated the exponent $\eta$ is between 1.5 and 2.8. \cite{Lipu15} investigated evolution of a thin accretion disc with a viscosity in proportion to a power-law function of the radial distance using the Green's functions for any outer boundary conditions. Just recently, \cite{rafikov16} found new similarity solutions for the evolution of a disc with non-zero torque and accretion at its center. 

A more elaborate  case is when the viscosity is a power-law function of the surface density and the radial distance, i.e. $\nu \propto \Sigma^a r^b$, for which the diffusion-type equation for the surface density becomes nonlinear and similarity or numerical methods are needed for solving the equation. For an infinite disc with conserved total angular momentum, the exponent of the accretion rate decay is found $\eta = 5/4$ for Thomson opacity, and, the exponent becomes $\eta =19/16$ for the Kramer's opacity \citep[e.g.,][]{filipov, cannizzo}. For a finite disc, however, \cite{lipu2000}  found that the accretion rate decreases much faster, i.e. $\eta = 10/3$ (Kramers's opacity) and $\eta =5/2$ (Thomson opacity). Although the generated turbulence within the disc is widely believed to be due to the magnetorotational instability,  none of the mentioned works on the evolution of the disc proposed their viscosity model based on physical arguments. A few authors, however, presented their models for the evolution of the discs using physically motivated forms for the viscosity. For example, \cite{lin87} derived analytical solution for the evolution of a disc with gravity-driven turbulence and the exponent of the accretion rate decay is found $\eta = 6/5$; however, for a disc with convectively driven turbulence the exponent is found $\eta = 15/14$ by \cite{lin82}.

Overall, the conclusions for the considered cases are qualitatively the same: the accretion rate and the total mass of the disc decrease with age of the system, however, rate of these decays depend on the input parameters such as the proposed form for the disc viscosity. 

Various mechanisms have been proposed for the angular momentum transport and turbulence generation within an accretion disc depending on the physical properties of the disc  \citep[for a review, e.g.,][]{Balbus2003}. At present the consensus is that turbulence can be driven by the magnetorotational instability which is shown to operate in the poorly ionized parts of a disc  \citep{Balbus91}, however, in the regions of a disc where none of the known sources of heat is able to ionize the gas  \citep{gam96}, it turns out that other mechanisms are needed for the angular momentum transport  \citep[e.g.,][]{lin87,Rice2003,rafikov2009,Clarke09,Cossins2010,Martin13,Dong16}. Gravity-driven turbulence is the mostly likely candidate, particularly at the outer parts of a PPD \citep[e.g.,][]{Mat2005,Kratter2008,rice2009,Voro2013,Rafikov15}. Although gravitational stability of a thin accretion disc is examined via a parameter, known as Toomre parameter \citep{Toomre64}, numerical simulations of self-gravitating accretion discs and physical arguments showed that thermal physics of a disc has also a vital role in the gravitational stability of an accretion disc  \citep[][]{gami,Cossins2010,Rice2003,rafikov2009,Cai2010,Rice2014,Tsukamoto15}.

Since thermal time-scale is much shorter than viscous time-scale, the  disc will be in thermal equilibrium over a longer period of time which then implies that there is a balance between heating and cooling rates. This constraint dictates a unique value for the stress parameter $\alpha$, and,  it enables us to construct steady-state  models for the self-gravitating accretion discs  \citep[][]{rafikov2009,Rafikov15}. This line of argument is consistent with the numerical simulations which show that a gravity-driven accretion disc settles into a state with a Toomre parameter close to its threshold and evolution of the disc will depend on the thermal properties of the disc.   With a Toomre parameter close to its threshold, if the disc losses generated heat due to the gravity-driven turbulence over a period of time faster than a few times of the dynamical time-scale, then the disc is subject to the fragmentation, however, in the opposite limit, the disc does not fragment and the disc is in a state which is known as a {\it gravitoturbulent} state. In other words, a gravitoturbulent state refers to a model for the self-gravitating accretion discs, in which the Toomre parameter is kept fixed around its critical value for the instability and the stress parameter $\alpha$, instead of being an input parameter, is determined self-consistently from the cooling rate with the assumption that the disc is in the thermal equilibrium. Under these circumstances, theoretical considerations in agreement with the numerical simulations show that the stress parameter can be written as $\alpha \simeq (\Omega t_{\rm cool})^{-1}$, where $\Omega$ is the Keplerian angular velocity and $t_{\rm cool}$ is the cooling time-scale  \citep[e.g.,][]{gami,Rafikov2005,rafikov2009}. According to the numerical simulations, there is a maximum value $\alpha_{\rm C}$ for the stress parameter so that once $\alpha$ becomes greater than $\alpha_{\rm C}$, the disc is susceptible  to the fragmentation  \citep[e.g.,][]{Rice2003,Rice2005}. It means that fragmentation of a disc depends on weather the disc is able to lose its internal thermal energy over a time-scale shorter than the dynamical time-scale.

Most of the previous analytical models for the gravitoturbulent discs are  restricted to the steady-state case where all the disc quantities are  assumed to be independent of time  \citep[][]{rafikov2009,Rafikov15}. In this regard, these models are similar to the standard accretion disc model, however, there are two main differences: (i) the stress parameter is determined from the cooling rate, (ii) Toomre parameter of the disc is  a fixed value close to its threshold. Note that some authors have already used only the second assumption for constructing their self-regulated disc models, however, they had relaxed the energy equation  \citep[e.g.,][]{Collin}.  However, a few authors have tried to investigate evolution of gravitoturbulent discs by extending the standard disc model to the time-dependent case, in which the disc settles into a thermal equilibrium state whereas its dynamical evolution is controlled by the gravity-driven turbulence.  For instance, \cite{rice2009} investigated time-dependent behavior  of a gravitoturbulent disc using a local approach for the gravity-driven viscosity, similar to the previous steady-state models \citep[e.g.,][]{rafikov2009}, which is a valid approximation so long as the mass of the disc is smaller than the central object  \citep[e.g.,][]{Balbus99,Lodato2004}. Although the spatial dependence of the physical quantities in their model is different from the steady model of \cite{rafikov2009}, possibly due to different model for the opacity, they found that the disc quickly attain a quasi-steady state. An interesting feature of their analysis is that the accretion rate is found to be largely independent of the radial distance.

In this work, we investigate evolution of a thin self-gravitating accretion disc with the gravity-driven turbulence using our new  analytical solutions. Numerical solutions of \cite{rice2009} for the evolution of a gravitoturbulent accretion disc exhibit an interesting feature which greatly helped us to obtain our analytical solutions. \cite{rice2009} showed that the accretion rate quickly becomes independent of the radial distance and it becomes solely a function of time. It is worth to note that many previous authors who studied evolution of the thin accretion discs, irrespective of true nature of the viscosity, have already mentioned that their numerical solutions correspond to the accretion rates independent of the spatial coordinates. This key finding motivated \cite{chambers} to construct fully analytical solutions for the evolution of an accretion disc in cases with and without central irradiation. However, \cite{chambers} implemented the usual $\alpha -$model for the disc viscosity, without paying attention to the possible mechanisms that may lead to the assumed form for the viscosity. Moreover, a fixed value for the opacity is used in \cite{chambers}, which it seems to be a great simplification. 

In this work, although we closely follow the mathematical approach of \cite{chambers} to obtain analytical solutions for the evolution of a self-gravitating disc, a noticeable advantage is that we do not prescribe viscosity in an {\it ad hoc} fashion, similar to the previous studies. Instead, as we mentioned earlier, in a gravitoturbulent disc model, the stress parameter is obtained self-consistently from the basic physics of the system. Furthermore, we consider optical thickness of the disc. At the early phase of disc evolution, the accretion rate is high and the disc becomes optically thick; however, as the system evolves toward the later phases, not only the accretion rate significantly reduces but also the optical depth becomes low. Consequently, at an  intermediate phase, the structure of a disc can be considered as an inner optically thick region and an outer optically thin part. In our model, we consider optically thick and thin discs separately, which are applicable to a gravity-driven disc at the early and final stages of its evolution. In the next section, we present general equations of our model. Then, time-dependent solutions for the optically thick and thin cases are obtained in sections 3 and 4, respectively. We conclude by a summary of our results in section 6.

\section{general formulation}
Our approach to construct a gravitoturbulent model is a direct generalization of \cite{rafikov2009} model to the time-dependent case. In agreement to the numerical simulations, we impose two key constraints to obtain the stress parameter $\alpha$ as a function of the disc physical quantities. First, the disc is assumed to evolve such that its Toomre parameter holds around a fixed value $Q_0$, i.e. 
\begin{equation}\label{eq:Q}
Q_0 = \frac{\Omega c_{\rm s}}{\pi G \Sigma},
\end{equation}
where $c_{\rm s}$ and $\Sigma$ are the sound speed and the surface density. Also, $\Omega$ is the Keplerian angular velocity, i.e. $\Omega = \sqrt{GM_{\star}/r^3}$, where $M_{\star}$ is the mass of the central star, and,  $r$ is the radial distance. Using ideal equations state for the gas, the sound speed is written as $c_{\rm s}=\sqrt{k_{\rm B}T/\mu m_{\rm H}}$, where $k_{\rm B}$, $T$, $\mu$, and, $m_{\rm H}$ are Boltzman constant, temperature, mean molecular weight, and the hydrogen mass, respectively. 

The second constraint comes from thermal physics of the disc which gives the stress parameter as a function of the cooling time-scale \cite[e.g.,][]{gami}, i.e.
\begin{equation}\label{eq:alpha-cool}
\alpha \simeq \frac{1}{\Omega t_{\rm cool}},
\end{equation}
where $t_{\rm cool}$ is the cooling time-scale. Assuming that the disc is in thermal equilibrium implies that
\begin{equation}\label{eq:tcool}
t_{\rm cool} = \frac{\Sigma c_{\rm s}^2}{\sigma T^4} f(\tau),
\end{equation}
where $\sigma$ is the Stephan-Blotzman constant. Here, the optical depth is denoted by $\tau = \kappa \Sigma$, where  the gas opacity, $\kappa$, is a complicated function of the density and the temperature. The function $f(\tau)$ is introduced to smoothly interpolates between optically thick regime ($\tau \gg 1$), and, optically thin case ($\tau \ll 1$). This function is written as $f(\tau ) = \tau + \tau^{-1}$.

Structure and evolution of PPDs are significantly affected by the adopted  functional form of the opacity as a function of the disc quantities. As long as  temperature of the disc is low the opacity is due to the dust particles, and so, $\kappa$ is approximated as a  power-law function of the temperature, i.e.
\begin{equation}\label{eq:kappa}
\kappa = \kappa_0 T^{\beta},
\end{equation}
where the exponent $\beta$ and the coefficient $\kappa_0$ depend on the temperature interval. For instance, when the opacity is due to the icy grains, we have $\beta =2$ and $\kappa_0 = 5\times 10^{-4}$ cm$^2$ g$^{-1}$ K$^{-2}$.

Using equations (\ref{eq:Q}), (\ref{eq:alpha-cool}), (\ref{eq:tcool}),  (\ref{eq:kappa}), and, the well-know equation for the viscosity, i.e. $\nu = \alpha c_{\rm s}^2 / \Omega $, the following expressions  for the opacity, the stress parameter, and, the viscosity coefficient are obtained:
 \begin{equation}\label{eq:tau}
\tau \approx \kappa_0 \Sigma^{2\beta + 1} \left ( \frac{\mu m_{\rm H}}{k_{\rm B}}\right )^{\beta} \left ( \frac{\pi G Q_0 }{\Omega}\right )^{2\beta},
 \end{equation}
\begin{equation}\label{eq:alpha}
\alpha = \zeta \frac{\sigma (\pi G Q_0 )^6}{f(\tau )} \left ( \frac{\mu m_{\rm H}}{k_{\rm B}}\right )^4 \frac{\Sigma^5}{\Omega^7},
\end{equation}
\begin{equation}\label{eq:nu}
\nu = \zeta  \frac{\sigma (\pi G Q_0 )^8}{f(\tau )} \left ( \frac{\mu m_{\rm H}}{k_{\rm B}}\right )^4 \frac{\Sigma^7}{\Omega^{10}},
\end{equation}
where parameter $\zeta \simeq 1$  is introduced because of using approximate relations for the stress parameter and the cooling time-scale.

The above equations are similar to what  have already been obtained by \cite{rafikov2009} for a steady-state gravitoturbulent disc, however, the above equations are valid even if the  disc quantities have both temporal and spatial dependence. Moreover, conservation of mass and the angular momentum lead to the following equation:
\begin{equation}\label{eq:evol}
\frac{\partial\Sigma}{\partial t} = \frac{3}{r} \frac{\partial}{\partial r} \left [ r^{1/2} \frac{\partial}{\partial r} \left ( \nu \Sigma r^{1/2} \right ) \right ].
\end{equation}

Upon substituting viscosity equation (\ref{eq:nu}) into  equation (\ref{eq:evol}), a partial differential equation for the surface density is obtained which can be solved numerically. Actually, \cite{rice2009} followed this approach, though their opacity model was slightly different from ours. Their numerical solutions exhibit an interesting feature, i.e. the disc evolves so   that the accretion rate quickly becomes independent of the radial distance, irrespective of the imposed initial conditions. Figures 4 and 9 of \cite{rice2009} clearly show that the accretion rate decreases with the age of system, however, this trend is independent of radius in the disc. Previous studies of thin disc evolution, based on the numerical integration of the above diffusion equation, have also found that neglecting variation of the accretion rate with radius in the disc is roughly a good approximation  for describing the behavior of the actual disks \citep[e.g.,][]{ruden86,morfill,garaud,kennedy}. Using this approximation,  \cite{step98} and \cite{chambers}  constructed their analytical solutions for the evolution of a viscous disc by assuming that the disc viscosity follows the standard "alpha" model \citep{SSmodel}. In their approach, the accretion rate only depends on time and is equal to the equilibrium accretion rate, so that
\begin{equation}\label{eq:main}
\dot{M} = 3\uppi \nu \Sigma.
\end{equation}

We also follow a similar approach to explore time-dependent behavior of a gravitoturbulent accretion disc. Two extreme cases are considered:  the entire disc is optically thick or the entire disc is optically thin. It will enable us to obtain analytical solutions for the evolution of the disc, however, optically thick solutions are applicable to the early phase of the disc evolution, but optically thin solutions are adequate for the final phases of disc evolution. In the next two sections, we obtain analytical solutions for these cases.

\section{Optically thick solutions}
In the optically thick case ($\tau\gg 1$), the function $f$ is approximated as $f(\tau) \simeq \tau $ and from equations (\ref{eq:kappa}), (\ref{eq:nu}), and (\ref{eq:main}), we obtain
\begin{equation}
\Sigma = \left(3\uppi C_{\rm thick}^{\nu} \right)^{-\frac{1}{7-2\beta}} \dot{M}^{\frac{1}{7-2\beta}} r^{\frac{3\beta -15}{7-2\beta}},
\end{equation}
where $C_{\rm thick}^{\nu}$ is 
\begin{equation}
C_{\rm thick}^{\nu}= \zeta  \sigma {\kappa_0}^{-1}  {(\uppi Q_0)^{-2\beta+8}} G^{-\beta+3} {(\mu m_{\rm H}/k_B)}^{-\beta+4} M_{\star}^{\beta-5}.
\end{equation}
We can normalize the accretion rate by the initial accretion rate, $\dot{M}_{0}$, and the radial distance by the initial radius of the disc outer edge, $s_0$. Thus, the above equation for the surface density is rewritten as
\begin{equation}\label{eq:density-thick}
\Sigma = \Sigma_{\rm f}  (\frac{\dot{M}}{\dot{M}_0}  )^{\frac{1}{7-2\beta}} (\frac{r}{s_0} )^{\frac{3\beta -15}{7-2\beta}},
\end{equation}
where
\begin{equation}
\Sigma_{\rm f}=\left(3\uppi C_{\rm thick}^{\nu} \right)^{-\frac{1}{7-2\beta}} \dot{M}_{0}^{\frac{1}{7-2\beta}} s_{0}^{\frac{3\beta -15}{7-2\beta}}.
\end{equation}

Using equations (\ref{eq:Q}), (\ref{eq:density-thick}), and the equation of state for the ideal gas, we can obtain temperature  as a function of the radial distance and the accretion rate:
\begin{equation}
T=T_{\rm f} (\frac{\dot{M}}{\dot{M}_0})^{\frac{2}{7-2\beta}} (\frac{r}{s_0})^{-\frac{9}{7-2\beta}},
\end{equation}
where
\begin{equation}
T_{\rm f} = (\uppi Q_0 )^2 G (\mu m_{\rm H} /k_{\rm B})  s_{0}^{3} M_{\star}^{-1} \Sigma_{\rm f}^{2}.
\end{equation}

Moreover, optical depth and the stress parameter are obtained from equations (\ref{eq:tau}) and (\ref{eq:alpha}) which are simplified to the following expressions:
\begin{equation}
\tau = \tau_{\rm f} (\frac{\dot{M}}{\dot{M}_0})^{\frac{2\beta +1}{7-2\beta}} (\frac{r}{s_0})^{-\frac{6\beta + 15}{7-2\beta}},
\end{equation}
\begin{equation}
\alpha_{\rm thick}=C_{\rm thick}^{\alpha} s_{0}^{-3\beta + \frac{21}{2}} \Sigma_{\rm f}^{4-2\beta} (\frac{\dot{M}}{\dot{M}_0})^{\frac{4-2\beta}{7-2\beta}} (\frac{r}{s_0})^{\frac{27}{2(7-2\beta)}},
\end{equation}
where
\begin{equation}
\tau_{\rm f}= \kappa_{0} (\mu m_{\rm H}/k_{\rm B})^{\beta} (\uppi Q_0 )^{2\beta } G^{\beta} M_{\star}^{-\beta} s_{0}^{3\beta} \Sigma_{\rm f}^{1+2\beta},
\end{equation}
\begin{equation}
C_{\rm thick}^{\alpha}= \zeta \sigma {\kappa_0}^{-1}  {(\uppi Q_0)^{-2\beta+6}} G^{-\beta+\frac{5}{2}}  {(\mu m_{\rm H}/k_B)}^{-\beta+4} M_{\star}^{\beta-\frac{7}{2}}.
\end{equation}

Using equation (\ref{eq:density-thick}) for the surface density, we now seek to calculate the total mass and the total angular momentum of the disc. The mass of the entire disc is obtained by  $M_{\rm d}=2\pi \int_{s_{\rm in}}^{s} r\Sigma dr$, where $s_{\rm in}$ is the inner radius of the disc. Thus, we obtain
\begin{displaymath}
M_{\rm d}=2\pi (\frac{ 7-2\beta }{1+\beta }) s_{0}^{2} \Sigma_{\rm f} (\frac{\dot{M}}{\dot{M}_0 })^{\frac{1}{7-2\beta}} 
\end{displaymath}
\begin{equation}\label{eq:Md-thick0}
\times [(\frac{s_{\rm in}}{s_0 })^{-\frac{1+\beta}{7-2\beta}} - (\frac{s}{s_0})^{-\frac{1+\beta}{7-2\beta}}].
\end{equation}

The total angular momentum is $L=2\pi \sqrt{G M_{\star}} \int_{s_{\rm in}}^{s} r^{3/2}  \Sigma dr $, and using equation (\ref{eq:density-thick}) the integral is calculated as
\begin{displaymath}
L=4\pi (\frac{7-2\beta}{4\beta -5}) \sqrt{GM_{\star}} s_{0}^{5/2} \Sigma_{\rm f} (\frac{\dot{M}}{\dot{M}_0 })^{\frac{1}{7-2\beta}} 
\end{displaymath}
\begin{equation}\label{eq:L-thick0}
\times [(\frac{s_{\rm in}}{s_0 })^{-\frac{4\beta -5}{2(7-2\beta)}} - (\frac{s}{s_0})^{-\frac{4\beta -5}{2(7-2\beta)}}].
\end{equation}

So far, our main equations have been presented without specifying range of the temperature, and consequently   a certain value for the exponent of opacity. Therefore,  from now on, we consider $\beta =2$ which corresponds to a case where opacity is due to the icy grains. \cite{rafikov2009} also explored properties of the steady-state gravitoturbulent discs for this particular case. Although we  investigate  evolution of a gravitoturbulent disc with $\beta =2$, one can easily explore properties of the solutions using the above equations for other values of $\beta$. We think, however, behavior of the solutions for other values of $\beta$ would be similar to what we present for $\beta =2$. 

If we set $\beta =2$, then equations (\ref{eq:Md-thick0}) and (\ref{eq:L-thick0}) give the following expressions for  mass of the disk and its total angular momentum:
\begin{equation}\label{eq:M}
M_{\rm d}=2\pi s_{0}^2 \Sigma_{\rm f} (\frac{\dot{M}}{\dot{M}_0 })^{\frac{1}{3}} [(\frac{s_{\rm in}}{s_0 })^{-1} - (\frac{s}{s_0})^{-1}],
\end{equation}
\begin{equation}\label{eq:L}
L=4\pi \sqrt{GM_{\star}} s_{0}^{5/2} \Sigma_{\rm f} (\frac{\dot{M}}{\dot{M}_0 })^{\frac{1}{3}} [(\frac{s_{\rm in}}{s_0 })^{-\frac{1}{2}} - (\frac{s}{s_0})^{-\frac{1}{2}}].
\end{equation}
By eliminating $\dot{M}$ between equations (\ref{eq:M}) and (\ref{eq:L}), we obtain
\begin{equation}
M_{\rm d}=\frac{L}{2\sqrt{GM_{\star}} s_{0}^{1/2}}\frac{(\frac{s_{\rm in}}{s_0 })^{-1} - (\frac{s}{s_0})^{-1}}{(\frac{s_{\rm in}}{s_0 })^{-\frac{1}{2}} - (\frac{s}{s_0})^{-\frac{1}{2}}}.
\end{equation}
One can easily simplify this equation and the outer edge radius is obtained as a function of the total mass, i.e.
\begin{equation}\label{eq:s-M}
\frac{s}{s_0}=[\frac{2}{\lambda}(\frac{M_{\rm d}}{M_{\rm 0d}})-(\frac{s_{\rm in}}{s_0})^{-\frac{1}{2}}]^{-2},
\end{equation}
where $M_{\rm 0d}$ is the initial mass of the disc, and, $\lambda$ is a dimensionless parameter: $\lambda = L/(M_{\rm 0d} \sqrt{s_{0}GM_{\star}})$. Considering the initial conditions, i.e. $M_{\rm d}(t=0)=M_{\rm 0d}$ and $s(t=0)=s_0$, equation (\ref{eq:s-M}) reduces to 
\begin{equation}\label{eq:lambda}
1=\frac{2}{\lambda} - (\frac{s_{\rm in}}{s_0})^{-\frac{1}{2}}.
\end{equation}

Since the total angular momentum of the disc is conserved, from equation (\ref{eq:L}) we obtain
\begin{equation}\label{eq:Mdot-M}
(\frac{\dot{M}}{\dot{M}_0})^{\frac{1}{3}} [(\frac{s_{\rm in}}{s_0})^{-\frac{1}{2}}-(\frac{s}{s_0})^{-\frac{1}{2}}]={\cal C},
\end{equation}
where ${\cal C}$ is a constant parameter. From initial conditions, this parameter becomes  ${\cal C}=(s_{\rm in}/s_0)^{-1/2} -1$. Using equations (\ref{eq:s-M}) and (\ref{eq:lambda}), therefore, equation (\ref{eq:Mdot-M}) is simplified to 
\begin{equation}\label{Mdot-1}
\dot{M}= \frac{(1-\lambda)^3}{(2-\lambda - \frac{M_{\rm d}}{M_{\rm 0d}})^3} \dot{M}_0 .
\end{equation}

It is evident that in the absence of any source or sink of mass for the disc, its total mass reduces mainly due to the accretion onto the central object. Thus, we have $dM_{\rm d}/dt = - \dot{M}$, which leads to the following differential equation:
\begin{equation}
\frac{dM_{\rm d}}{dt}= - \frac{(1-\lambda)^3}{(2-\lambda - \frac{M_{\rm d}}{M_{\rm 0d}})^3} \dot{M}_0 .
\end{equation} 
Fortunately, this equation is analytically integrable and its solution is
\begin{equation}\label{eq:Md}
\frac{M_{\rm d} (t)}{M_{\rm 0d}} = 2-\lambda - (1-\lambda ) (1+\frac{t}{t_0} )^{\frac{1}{4}}, 
\end{equation}
where
\begin{equation}
t_0 = \frac{1-\lambda}{4} (\frac{M_{\rm 0d}}{\dot{M}_0}).
\end{equation}
Upon substituting equation (\ref{eq:Md}) into equation (\ref{eq:s-M}), the outer edge radius as a function of time is obtained, i.e.
\begin{equation}\label{eq:outer-thick}
\frac{s(t)}{s_0} =\frac{\lambda^2}{4 (1-\lambda )^2} [\frac{2-\lambda}{2(1-\lambda )} - (1+\frac{t}{t_0})^{1/4}]^{-2}.
\end{equation}
Now, if we substitute equation (\ref{eq:Md}) into equation (\ref{Mdot-1}), the accretion rate becomes
\begin{equation}
\dot{M}= \frac{\dot{M}_0}{(1+\frac{t}{t_0})^{3/4}},
\end{equation}
and the rest of disc quantities become
\begin{displaymath}
\Sigma (r, t) = 3.1 (\frac{M_{\star}}{M_{\odot}})(\frac{\dot{M}_0}{10^{-8} {\rm M}_{\odot}/{\rm yr}} )^{\frac{1}{3}} (\frac{s_0}{100 {\rm AU}})^{-3}
\end{displaymath}
\begin{equation}
\times (1+\frac{t}{t_0})^{-\frac{1}{4}} (\frac{r}{s_0})^{-3} \hspace{0.5cm} {\rm g} {\rm cm}^{-2},
\end{equation}
\begin{displaymath}
T (r, t) = 0.268 (\frac{M_{\star}}{M_{\odot}}) (\frac{\dot{M}_0}{10^{-8} {\rm M}_{\odot}/{\rm yr}} )^{\frac{2}{3}} (\frac{s_0}{100 {\rm AU}})^{-3}
\end{displaymath}
\begin{equation}\label{eq:Temp-thick}
\times (1+\frac{t}{t_0})^{-\frac{1}{2}} (\frac{r}{s_0})^{-3} \hspace{0.5cm} {\rm K},
\end{equation}
\begin{displaymath}
\tau (r, t) = 1.1\times 10^{-4} (\frac{M_{\star}}{M_{\odot}})^{3}(\frac{\dot{M}_0}{10^{-8} {\rm M}_{\odot}/{\rm yr}} )^{\frac{5}{3}} (\frac{s_0}{100 {\rm AU}})^{-9}
\end{displaymath}
\begin{equation}\label{eq:tau-thick}
\times (1+\frac{t}{t_0})^{-\frac{5}{4}} (\frac{r}{s_0})^{-9}.
\end{equation}

The scaling factor $10^{-8}$ M$_{\odot}$/yr for the accretion rate is typical of classical T Tauri stars. 
It is also worth mentioning  that the above solutions are valid so long as the optical depth stays larger than one. Moreover, the optical depth decreases as we move out toward the outer edge of the disc which means that $\tau$ reaches to its minimum at the outer edge $s$. Therefore, we explore evolution of a disc up to when the optical depth at the other edge becomes around unity, i.e. $\tau (r=s(t_{\rm life}), t=t_{\rm life}) =1$. Here, we assumed that the optical depth at the outer edge drops to one after time $t_{\rm life}$ which implies that the above solutions are valid for the times less than $t_{\rm life}$. Using equations (\ref{eq:outer-thick}) and (\ref{eq:tau-thick}), one can easily estimate $t_{\rm life}$. We obtain
\begin{displaymath}
28.8 (\frac{M_{\star}}{M_{\odot}})^{3} (\frac{\dot{M}_0}{10^{-8} {\rm M}_{\odot}/{\rm yr}} )^{\frac{5}{3}} (\frac{s_0}{100 {\rm AU}})^{-9} (\frac{1-\lambda}{\lambda})^{18}
\end{displaymath}
\begin{equation}\label{eq:thick-valid}
(1+\frac{t_{\rm life}}{t_0})^{-\frac{5}{4}} [\frac{2-\lambda}{2(1-\lambda)}-(1+\frac{t_{\rm life}}{t_0})^{\frac{1}{4}}]^{18}=1.
\end{equation}
\begin{figure}
\includegraphics[scale=0.5]{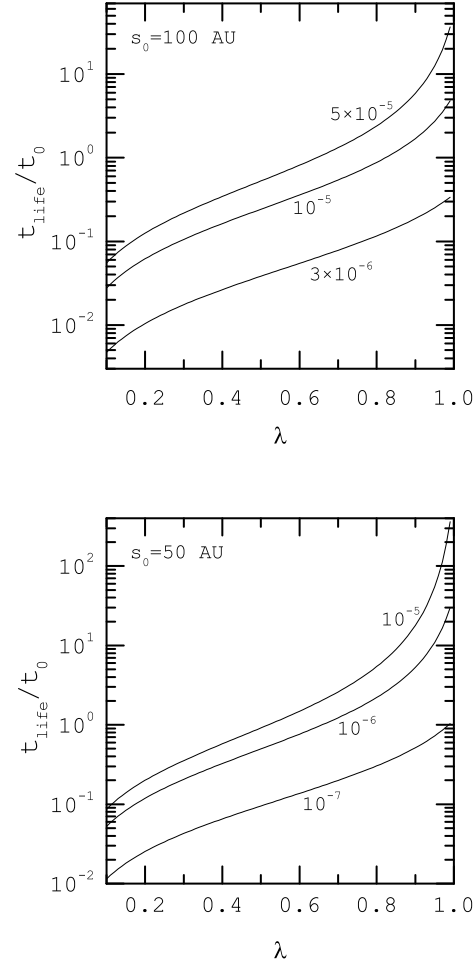}
\caption{The profile of $t_{\rm life}/t_0$ as a function of the parameter $\lambda$ for different initial accretion rates, and, different initial size of the disc: $s_0 = 100$ AU (top) and $s_0 =50$ AU (bottom). Mass of the central star is $M_{\star}=M_{\odot}$. It is important to note that the time-scale $t_0$ depends on the initial accretion rate, initial mass of the disc, and, the parameter $\lambda$. }\label{fig:f1}
\end{figure}
This equation can be solved numerically for given parameters $\dot{M}_0$, $s_0$, $\lambda$, and the mass of the central star, however, we found that equation (\ref{eq:thick-valid}) has no real solution once the initial accretion rate drops to values less than a critical value for a given initial outer radius. Figure \ref{fig:f1} shows profile of $t_{\rm life}/t_0$ as a function of the parameter $\lambda$ for different initial accretion rate and the initial outer edge of the disc. Mass of the central star is assumed to be one solar mass. Figure \ref{fig:f1} (top) displays cases with a fixed initial size of the disc, i.e. $s_0 =100$ AU, and different values for $\dot{M}_0$. Each curve is labeled by the corresponding initial accretion rate in solar mass per year. For a given parameter $\lambda$, our solutions become valid over a longer period of time as the initial accretion rate becomes larger. For this particular set of the input parameters, however, the time-scale $t_{\rm life}$ becomes negligible once the initial accretion rate drops to values smaller than around $3\times 10^{-6}$ ${\rm M}_{\odot}/{\rm yr}$. In other words, the above optically thick solutions are not appropriate for a disc with $s_0 = 100$ AU, if the initial accretion rate is less than $10^{-5}$ ${\rm M}_{\odot}/{\rm yr}$, irrespective of parameter $\lambda$. According to equation (\ref{eq:lambda}), we note that parameter $\lambda$ quantifies the ratio of inner and the outer radii of a disc. If the radial extension of a disc becomes smaller (i.e., the outer radius is closer to the inner radius), then the parameter $\lambda$ becomes closer to unity. Figure \ref{fig:f1} (top) shows that the ratio $t_{\rm life}/t_0$ becomes larger, if the parameter $\lambda$ tends to unity which means that optically thick solutions are able to describe the evolution of a smaller disc during a longer period of time comparing to a similar, but larger disc. In Fig. \ref{fig:f1} (bottom), we consider cases with a smaller initial outer radius, i.e. $s_0 =50$ AU. The overall behavior of $t_{\rm life}/t_0$ as a function of $\lambda$ is similar to the top plot, however, the solutions become invalid for the initial accretion rate smaller than $ 10^{-7}$ ${\rm M}_{\odot}/{\rm yr}$. In other words, optically thick solutions are applicable over a longer period of time for a disc with a smaller radial extension comparing to a disc with a larger size.

We now turn to the question that how location of the snow line evolves over the time based on our optically thick solutions. Snow line in a PPD is defined by the radial distance at which the temperature is around $T_{\rm snow}\simeq 170$ K. It means that interior to the snow line radius, the temperature is larger than 170 K and icy planets can not form, however, temperature  beyond the snow line is smaller than 170 K and formation of icy planets is possible. Using equation (\ref{eq:Temp-thick}) for the temperature profile, we can determine location of the snow line, i.e.  
\begin{displaymath}
\frac{r_{\rm snow}(t)}{s_0}= (\frac{0.268 {\rm K}}{T_{\rm snow}})^{\frac{1}{3}} (\frac{M_{\star}}{M_{\odot}})^{\frac{1}{3}} (\frac{\dot{M}_0}{10^{-8} {\rm M}_{\odot}/{\rm yr}})^{\frac{2}{9}} 
\end{displaymath}
\begin{equation}\label{eq:snow-thick}
\times (\frac{s_0}{100 {\rm AU}})^{-1} (1+\frac{t}{t_0})^{-\frac{1}{6}}.
\end{equation}
This equations shows that location of the snow line migrates inward as the disc evolves: $r_{\rm snow} \propto t^{-1/6}$. Moreover, location of the snow line is proportional to $M_{\star}^{1/3}$, which suggests that in a PPD with a massive central protostar, $r_{\rm snow}$ lies at a larger distance  comparing to a low-mass protostar. We can now consider a fiducial model with $M_{\star}=M_{\odot}$, $s_0 = 50$ AU, and, $\dot{M}_0 = 10^{-6}$ M$_{\odot}$/yr. Assuming the inner radius as $s_{\rm in}=10$ AU, and the initial mass of the disc as $M_{\rm 0d}=0.01$ M$_{\odot}$, we then obtain $\lambda \simeq 0.6$ and $t_{0}\simeq 10^3$ yr. For these input parameters, Fig. \ref{fig:f1} implies that $t_{\rm life}\simeq t_0$. From equation (\ref{eq:snow-thick}), the snow line as a function of time becomes $r_{\rm snow} \simeq 32 (1+t/t_0 )^{-1/6}$ AU. It implies that the snow line begins at radius 32 AU, however, its location gradually moves inward  and by the time that our solution is valid (i.e., $t_{\rm life} =10^3$ yr) the snow line reaches to around 28 AU.  We can not follow evolution of the snow line beyond $10^3$ yr using optically thick solutions because the disc becomes optically thin and the corresponding solutions should be used. However, if we had considered an initially more massive disc, we would be able to follow evolution of the snow line over a longer period of time.

\section{Optically thin solutions}
The optically thin regime corresponds to a disc with depleted gas component which implies that the disc has a very low surface density. One can expect this situation is realized for very low accretion rates. When the disc is optically thin ($\tau\ll 1$), the function $f$ is approximated as $f(\tau)\simeq \tau^{-1}$. Disc quantities are obtained following the approach that we used in the optically thick regime. The surface density becomes 
\begin{equation}
\Sigma = \Sigma_{\rm f} (\frac{\dot{M}}{\dot{M}_0})^{\frac{1}{2\beta +9}} (\frac{r}{s_0})^{-\frac{3\beta +15}{2\beta +9}} ,
\end{equation}
where
\begin{equation}
\Sigma_{\rm f} = \left(3\pi C_{\rm thin}^{\nu} \right)^{-\frac{1}{2\beta +9}} \dot{M}_{0}^{\frac{1}{2\beta +9}} s_{0}^{-\frac{3\beta +15}{2\beta +9}},
\end{equation}
and
\begin{equation}
C_{\rm thin}^\nu = \zeta \sigma \kappa_0 (\uppi Q_0 )^{2\beta+8} G^{\beta + 3} (\mu m_{\rm H} /k_{B})^{\beta +4} M_{\star}^{-\beta - 5}.
\end{equation}

Thus, the total mass and the total angular momentum of the disc becomes
\begin{displaymath}
M_{\rm d} =2\uppi (\frac{2\beta +9}{\beta +3}) s_{0}^{2} \Sigma_{\rm f} (\frac{\dot{M}}{\dot{M}_{0}})^{\frac{1}{2\beta +9}}
\end{displaymath}
\begin{equation}
\times [(\frac{s}{s_0})^{\frac{\beta +3}{2\beta +9}} - (\frac{s_{\rm in}}{s_0})^{\frac{\beta +3}{2\beta +9}} ], 
\end{equation}
\begin{displaymath}
L =2\uppi (\frac{4\beta +18}{4\beta +15})\sqrt{GM_{\star}} s_{0}^{5/2} \Sigma_{\rm f} (\frac{\dot{M}}{\dot{M}_{0}})^{\frac{1}{2\beta +9}}
\end{displaymath}
\begin{equation}
\times [(\frac{s}{s_0})^{\frac{4\beta +15}{4\beta +18}} - (\frac{s_{\rm in}}{s_0})^{\frac{4\beta +15}{4\beta +18}} ].
\end{equation}
As before, we limit our calculations to a particular case with $\beta =2$. Therefore, the above equations for the mass and the angular momentum of the disc are simplified to the following expressions:
\begin{equation}
M_{\rm d} =\frac{26}{5}\uppi  s_{0}^{2} \Sigma_{\rm f} (\frac{\dot{M}}{\dot{M}_{0}})^{1/13} [(\frac{s}{s_0})^{5/13} - (\frac{s_{\rm in}}{s_0})^{5/13} ] ,
\end{equation}
\begin{equation}
L =\frac{52}{23}\pi \sqrt{GM_{\star}} s_{0}^{5/2} \Sigma_{\rm f} (\frac{\dot{M}}{\dot{M}_{0}})^{1/13} [(\frac{s}{s_0})^{23/26} - (\frac{s_{\rm in}}{s_0})^{23/26} ] .
\end{equation}

Here, for simplicity, the inner radius of the disc is assumed to be much smaller than the outer edge of the disc. We note that, however, one can easily relax this simplification and keep a non-zero value for $s_{\rm in}$. If we set $s_{\rm in} =0$, then we obtain
\begin{equation}
M_{\rm d} =  \frac{23 s_{0}^{-1/2} L}{10\sqrt{GM_{\star}}} (\frac{s}{s_0})^{-\frac{1}{2}}
\end{equation}
If the initial mass of the disc is denoted by $M_{\rm 0d}$, the above equation implies that
\begin{equation}
M_{\rm 0d} =  \frac{23 s_{0}^{-1/2} L}{10\sqrt{GM_{\star}}}.
\end{equation}
Thus,
\begin{equation}\label{eq:thick-MdM0}
M_{\rm d} = M_{\rm 0d} (\frac{s}{s_0})^{-\frac{1}{2}}.
\end{equation}
Moreover, conservation of the total angular momentum implies that
\begin{equation}\label{eq:thick-Ms}
(\frac{\dot{M}}{\dot{M}_0 })^{\frac{1}{13}} (\frac{s}{s_0})^{\frac{23}{26}}=1.
\end{equation}
By illuminating $s$ between equations (\ref{eq:thick-MdM0}) and (\ref{eq:thick-Ms}), we obtain
\begin{equation}\label{eq:Mdot-thick}
\dot{M} = \dot{M}_0 (\frac{M_{\rm d}}{M_0})^{23}.
\end{equation}

Since we have $dM_{\rm d}/dt = -\dot{M}$, then total mass of the disc as a function of time becomes
\begin{equation}\label{eq:mass-thick}
M_{\rm d}(t) =\frac{M_{\rm 0d}}{(1+\frac{t}{t_0})^{1/22}}
\end{equation}
where $t_0 = (1/22)(M_{\rm 0d}/\dot{M}_0 )$ is the decay time-scale. The value of this characteristic decay time-scale depends on the initial accretion rate and the initial total mass of the disc. Upon substituting equation (\ref{eq:mass-thick}) into equation (\ref{eq:Mdot-thick}), the accretion rate is obtained, i.e.
\begin{equation}
\dot{M} = \frac{\dot{M}_0 }{(1+\frac{t}{t_0})^{23/22}}.
\end{equation}
Furthermore, the outer edge of the disc increases with time as
\begin{equation}
s(t) = s_0 (1+\frac{t}{t_0})^{1/11}.
\end{equation}
We thereby arrive at the following relations for  the rest of disc quantities:
\begin{displaymath}
\Sigma (r,t) = 12.5 (\frac{M_{\star}}{M_{\odot}})^{\frac{7}{13}} (\frac{\dot{M}_0}{10^{-8} {\rm M}_{\odot}/{\rm yr}})^{\frac{1}{13}} (\frac{s_0}{100 {\rm AU}})^{-\frac{21}{13}}
\end{displaymath}
\begin{equation}
\times (1+\frac{t}{t_0})^{-\frac{23}{286}} (\frac{r}{s_0})^{-\frac{21}{13}} \hspace{0.5cm} {\rm g} {\rm cm}^{-2},
\end{equation}

\begin{displaymath}
T (r,t) = 4.42 (\frac{M_{\star}}{M_{\odot}})^{\frac{1}{13}} (\frac{\dot{M}_0}{10^{-8} {\rm M}_{\odot}/{\rm yr}})^{\frac{2}{13}} (\frac{s_0}{100 {\rm AU}})^{-\frac{3}{13}}
\end{displaymath}
\begin{equation}
\times (1+\frac{t}{t_0})^{-\frac{23}{143}} (\frac{r}{s_0})^{-\frac{3}{13}} \hspace{0.5cm} {\rm K},
\end{equation}

\begin{displaymath}
\tau (r,t) = 0.12 (\frac{M_{\star}}{M_{\odot}})^{\frac{9}{13}} (\frac{\dot{M}_0}{10^{-8} {\rm M}_{\odot}/{\rm yr}})^{\frac{5}{13}} (\frac{s_0}{100 {\rm AU}})^{-\frac{27}{13}}
\end{displaymath}
\begin{equation}
\times (1+\frac{t}{t_0})^{-\frac{115}{286}} (\frac{r}{s_0})^{-\frac{27}{13}},
\end{equation}
\begin{displaymath}
\alpha (r,t) = 0.006 (\frac{M_{\star}}{M_{\odot}})^{-\frac{3}{26}} (\frac{\dot{M}_0}{10^{-8} {\rm M}_{\odot}/{\rm yr}})^{\frac{10}{13}} (\frac{s_0}{100 {\rm AU}})^{\frac{9}{26}}
\end{displaymath}
\begin{equation}\label{eq:alpha-thin}
\times (1+\frac{t}{t_0})^{-\frac{115}{143}} (\frac{r}{s_0})^{\frac{9}{26}}.
\end{equation}

The radial scale of the stress parameter $\alpha$ shows that it increases with the distance, however, as the disc becomes older, the stress parameter $\alpha$ decreases with time  as $t^{-0.8}$. When the stress parameter reaches to its critical value $\alpha_{\rm c}$, the disc is subject to fragmentation. Although there are intense debates on the critical value $\alpha_{\rm c}$,  numerical simulations of the self-gravitating discs show that the critical value is $\alpha_c \simeq 0.06$. Equation (\ref{eq:alpha-thin}) enables us to determine the fragmenation radius $r_{\rm frag}$ as a function of time. This equation, however, implies that the stress parameter is smaller than its critical value over a large radial extend which means the disc is stable subject to the gravitational instability. 

One should note that optically thin solutions are generally applicable to the outer parts of a disc. These regions, however, are subject to the irradiation of the central star which has not been considered. Following \cite{chambers} who included stellar irradiation, we think, it would be straightforward to generalize the above optically thin solutions by considering irradiation of the central star. Steady-sate models of the gravitoturbulent discs, however, imply that irradiated regions are strongly affected by the incident radiation instead of the generated heat due to the turbulence.  

\section{conclusions}
The main impetus for our work is to investigate evolution of a gravitoturbulent PPD using analytical solutions. In contrast to the steady-state models, however, there are relatively few attempts to investigate time-dependent behavior of a gravitoturbulent PPD. To our knowledge, there is in the literature no analytical solutions for the evolution of a gravitoturbulent PPD. We presented analytical solutions for the time-dependent behavior of a gravitoturbulent PPD in both optically thick and thin regimes, however, our solutions are by no means exact.  We applied the following assumptions which enabled us to obtain the present solutions for the evolution of a disc in a gravitoturbulent state:

 (i) Similar to the standard disc model \citep{SSmodel}, the generated heat due to the turbulence is assumed to radiate out of the system immediately after generation. The disc is geometrically thin and in the vertical direction is in hydrostatic equilibrium. 
 
 (ii) Gravitational instability is assumed to be the dominant mechanism of the angular momentum transport. Although gravity-driven turbulence is intrinsically a non-local transport mechanism, a local description based on an effective viscosity is used in constructing the model which is a reasonable approximation so long as the mass of the disc is much smaller than the mass of the central object \citep{Balbus99,Lodato2004}.
 
 (iii) Although the gravitational instability is the main mechanism of the turbulence, the disc is not in a fragmentating state. In agreement with previous numerical simulations, however, we assume that the disc evolves so that it hovers  near the edge of gravitational instability which implies that Toomre parameter to be close to its threshold. This new constraint enabled us to obtain coefficient of the turbulent viscosity by assuming that the disc is in thermal equilibrium. The disc remains in the gravitoturbulent state as it evolves with time.
 
 (iv) The above assumptions have already been used by some authors in their attempts for constructing steady-state or time-dependent gravitoturbulent disc models. But in agreement to the earlier numerical studies, we implemented a key assumption which enabled us to obtain analytical time-dependent solutions. This reasonable approximation states that with the evolution of the disc, the accretion rate has only temporal dependence.

 Although establishing robust estimates is hampered by uncertainties in our input parameters, temporal dependence of disc quantities in a gravitoturbulent state are obtained analytically.  Our new analytical solutions for the time-dependent structure of a gravitoturbulent PPD exhibit the following common main features.
 
Not only the accretion rate but also the total mass of the disc decrease with the age of the system. Most notably decay of the accretion rate with time is found as a power-law function of time with an exponent $\eta$ equal to -0.75 and -1.04 for the  optically thick and thin cases, respectively. This exponent $\eta$ of the accretion rate decay has already been obtained by the previous authors under different circumstances \citep[e.g.,][]{lynden74,filipov,cannizzo,king98,tanaka2011,Lipu15}. These theoretical studies found that  the exponent $\eta$ is between 1 and 3. Despite of uncertainties on the adopted observational approach, however, \cite{hartmann98} found that the exponent $\eta$ in the T Tauri stars is between 1.5 and 2.8. Although these values are not very well constrainted, a value around $\eta\simeq 1.5$ is widely used by the community. In a gravitoturbulent disc, nevertheless, the exponent $\eta$ is close to the lower limit of the so far   obtained values for this exponent. This finding implies that decay of the accretion rate with time in a gravitoturbulent disc is slower than the other models.  
 
Our treatment for modeling a gravitoturbulent PPD showed a monotonic decrease of the snow line radius with time as a power-law function with the exponent -1/6.   
 Given the importance of the snow line location in the theories of planet formation, many authors have already determined the snow line radius under a variety of assumptions. While most previous studies focused on the steady-state models for determining location of the snow line, there are recent observational evidences  and theoretical arguments which put forward time-evolution of the snow line \citep[e.g.,][]{Zhang13,Piso15}. In this regard, our study is a theoretical analysis  following previous studies, but the present work addresses time-evolution of the snow line in a gravitoturbulent disc. Althgough most of the earlier studies on the location of the snow line are restricted to the discs in isolation, some authors investigated role of enviromental effects in determining location of the snow line.  \cite{Jin2015}, for instance, investigated time-evolution of the snow line in a disc embedded in a collapsing progenitor cloud core. They found that the snow line radius gradually increases due to influx onto the disc from the collapse of the cloud, reaches  a maximum, and then decreases with time (also see, \cite{Jin2010}). However, some authors studied time-evolution of the snow line in an isolated disc. While a fully turbulent disc model predicts that the snow line is too close to the central star, \cite{Martin13} showed that this unsatisfactory aspect can be resolved if a dead zone is considered (also see, \cite{Martin12}). Our solutions can be used for determining CO snow line as well. For instance, \cite{Martin14} determined location of the CO snow line using a time-dependent models with considering dead zone. They found that current CO snow line radius in our solar system is not compatible with a fully turbulent disc model and inclusion of the dead zone can resolve this problem.     
 
 The present analytical solutions can be used in other studies where evolution of the gaseous disc  is input of the model.  For instance, \cite{Pudritz} presented a detailed analysis for the dynamics of dust particles in a PPD using analytical solutions of \cite{chambers} for the gas component. This approach is justified as long as surface density of the dust particles is much smaller than the gas surface density. Under this condition, evolution of the gas component is independent of the dust particles. This scenario changes if the effect of dust particles on the ionization level is considered. When magnetorotational instability is the main mechanism of the turbulence, ionization level plays a vital role and affects strongly the generated turbulence, and thereby, dynamical structure of the disc. Under these circumstances, dust particles may indirectly affect dynamics of the gas component. However, these complicated aspects, though are very important, can be neglected as a first approximation in analyzing a PPD consisting of dust and gas components. As we mentioned earlier the analysis of \cite{Pudritz} is based on the solutions of \cite{chambers}, however, our analytical solution is a good starting point for investigating dynamics of dust particles in a gravitoturbulent PPD.

\acknowledgments
We are grateful to referee for his/her constructive report which improved the quality of this paper. 

\bibliographystyle{apj}
\bibliography{reference}

\end{document}